\documentclass[aps,prl,groupedaddress,twocolumn,showkeys,showpacs]{revtex4-1}
\usepackage{epsfig}
\usepackage{color}

\begin{document}


\title{Qualitative model of high-$T_c$ superconductivity}


\author{Nikolai A. Zarkevich}
\email[]{zarkev@ameslab.gov}
\affiliation{Ames Laboratory, U.S. Department of Energy, Ames IA 50011, USA}


\date{\today}

\begin{abstract}
We suggest a qualitative model of a high-$T_c$ superconductor, 
based on considerations of thermo\-dynamics of phase transitions. 
As an example, we consider the Mott transition and classify 5 solid phases around it. 
In our model, a combined electronic and structural instability
 causes segregation into either neutral or charged phases. 
A charged precipitate with a quantized electric charge is 
a collective excitation of electrons, stabilized by a collective athermal displacement of ions;  
this local variation of the charge density, accompanied by a local lattice deformation, can behave as a quasi\-particle. 
A condensate of charged bosonic quasi\-particles is responsible for the superconductivity. 
\end{abstract}


\maketitle

\section{\label{Introduction}Introduction}
Materials on the edge of their stability can have enormous lattice response to a perturbing field.  A responsive lattice is necessary for such phenomena as, for example, a giant caloric effect or a high-$T_c$ superconductivity. 

{\par}
A first-order phase transformation happens between two phases of unequal density.  Phases can coexist  
at equal pressure $P$, temperature $T$, and chemical potential $\mu$.  
Volume $V$ [$\mbox{\AA}^3$ per formula unit (f.u.)], inter\-atomic distances $d_a$ [\AA], density $\rho$ [g/cm$^3$], 
and local electronic [$\mbox{\AA}^{-3}$] and charge [$e^-/\mbox{\AA}^3$] densities  
are discontinuous at a 1st-order phase transition. 
Intermediate structures between the two phases are unstable; their segregation into the stable phases lowers  the total Gibbs free energy $G$. 

{\par}
	In our model, we consider a high-$T_c$ superconductor as a phase-segregated material. 
In this neutral material both phases are charged, but the volume of one phase is much smaller than that of the other. 
The charged phase with small volume has a high charge density,  
 and the Coulomb repulsion fractures it into tiny precipitates, which behave as quasi\-particles, with a quantized electric charge.  In conventional superconductors, such quasi\-particles are known as the Cooper pairs, composed by a collective motion of electrons and a lattice deformation, which binds an even number of  electrons (fermions) into one charged quasi\-particle (boson). 

{\par}
	Conventional superconductivity \cite{Onnes1913,Abrikosov2003,JETP35p1558y1959} happens due to electron-phonon coupling \cite{GinsburgLandau1950,PR106p162p1175y1957,JETP34p58p73y1958,JETP34p66y1958,PRB93p054517y2016}. 
More generally, superconductivity (both conventional and unconventional) happens due to coupling between a collective electronic excitation and a lattice response to it, which is a collective athermal displacement of atoms and ions.
Lattice deformations are responsible for coupling fermions (electrons) into bosons, and a Bose-Einstein condensate \cite{Bose1924} of charged quasi\-particles is responsible for the superconductivity. A larger lattice response  can result in a larger critical temperature $T_c$. 
Hence, a guided search for high-$T_c$ superconductors starts with a study of electronic and lattice instabilities. 

{\par}
	A high-$T_c$ superconductivity occurs around an instability.  Electronic and structural instabilities result in phase transformations.  One of them is the Mott transition \cite{Mott1949}.

{\par}
	The Mott transition is electronic by nature.  It happens due to a change of electronic structure, accompanied by a change in interatomic interactions, which drive atoms to their new equilibrium positions, thus relaxing interatomic distances $d_a$, volume $V$, and density $\rho$. 
At the same external stress,  temperature, and composition, 
two different electronic states have equilibrium at different lattice constants.  
Any intermediate crystal structures between those two terminal equilibria are not stable: 
they are destined to transform.  What is the speed of electronic and structural transformations?

{\par}
	Electromagnetic interactions, including those between electrons, propagate with the speed of light 
$c_{light} \approx 3 \! \times \! 10^8\,$m/s.  
Fermi velocity of conductive electrons in a metal is $v_F \sim 10^6\,$m/s  (e.g., 1570 km/s in copper). 
Lattice vibrations (phonons) propagate with the speed of sound $v_{sound} \sim 10^3\,$m/s 
(e.g., 4760 m/s for longitudinal
waves in an annealed copper \cite{CRChandbook2008} at room $T$).

In contrast, drift velocity $u_D = J / q n_q $ of carriers with charge $q$ and concentration $n_q$ in a conductor with current density $J = I/S$ is very modest.  
For example, in a copper wire ($n_q=8.5 \times 10^{28}\,\mbox{m}^{-3}$) with a cross section $S=1\, \mbox{mm}^2$ at constant direct current $I=1$A, drift velocity of electrons 
constitutes only 
$7 \times 10^{-5}\,$m/s. 

{\par}
	In a superconductor, each charged bosonic quasi-particle is accompanied (and held together) by a local lattice deformation, which follows its drift.  From the other hand, thermal atomic motion disturbs such a local lattice deformation, which acts as a ``glue'' for the quasi-particle.  Without such a ``glue'', those quasi-particles can no longer exist, and superconductivity is destroyed above the critical temperature $T_c$.   
A larger lattice response 
results in a higher $T_c$.  
In general, the lattice response is the largest near phase transitions.  
The quantum critical point (QCP) is an example of a phase boundary at 0K. 

{\par}
Below we propose a model of superconductivity in the vicinity of a phase boundary. 
This model can help in a guided search for novel high-$T_c$ superconductors.

\begin{figure}[t]
\includegraphics[width=75mm]{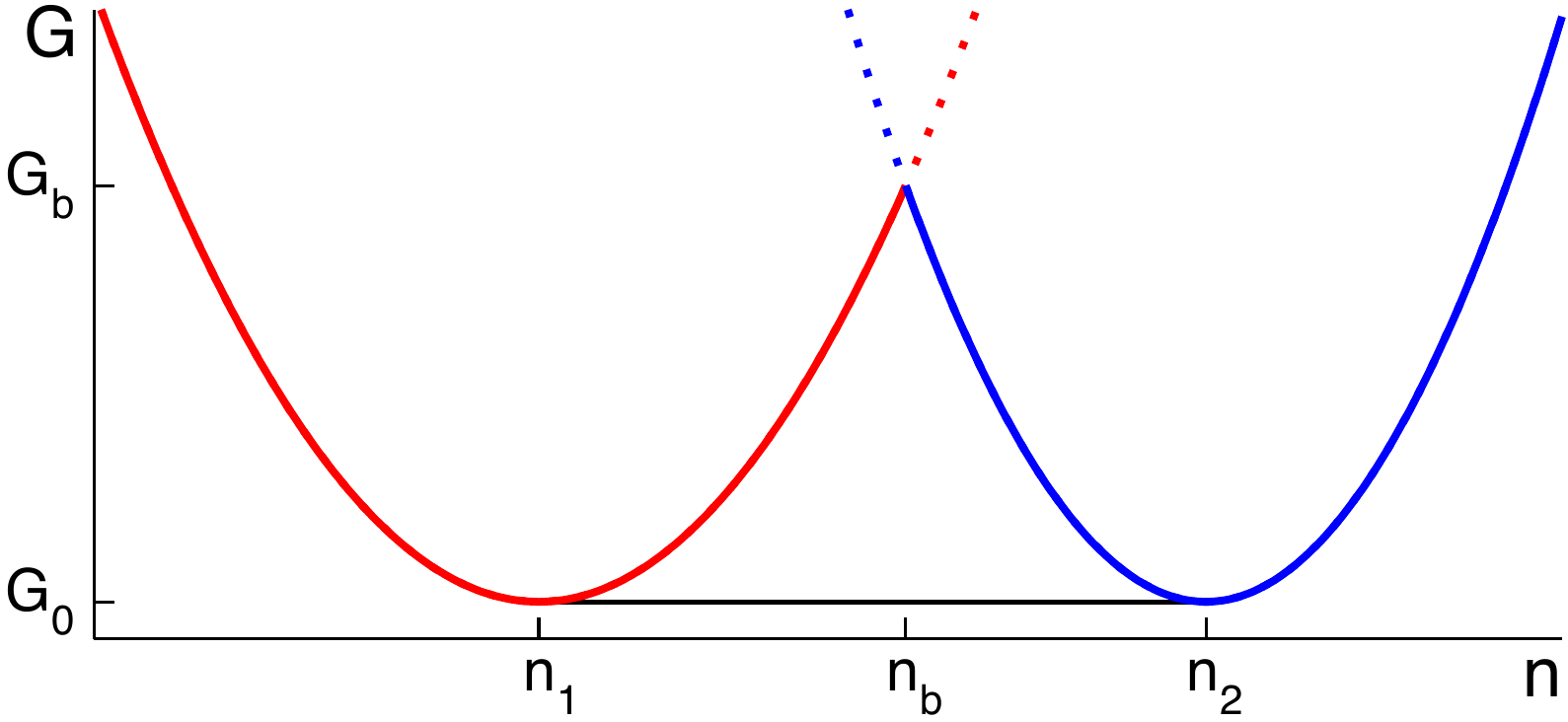}
\caption{\label{FigGn} Gibbs free energy $G$ of two phases (1 and 2) versus average electronic density $n$.  Its change $\Delta n$ can be comprehended as a level of doping.  The tangent (black line) is horizontal, if two segregated phases are at a thermal equilibrium. 
}
\end{figure}

\section{\label{Model}Theoretical Model} 
{\par} 
Let us consider a bulk solid with an instability, resulting in phase transformations. 
Such transformations can be driven by changing temperature $T$, pressure $P$,  composition $c$ (and hence chemical potential $\mu$), average electronic density $n$, or the level of doping $\Delta n$. 

\subsection{\label{Charge}Instability in electronic density}

{\par} 
Let us expand the Gibbs free energy $G(T,P,c,n)$ at fixed $\{T,P,c\}$ in a Taylor series around its minimum
 in each phase  $i=\{1,2\}$:
\begin{equation}
\label{eqG1}
  G_i (n) = G_i^{(0)} + (\partial^2 G_i / \partial n^2) (n-n_i)^2 + O(n-n_i)^3. 
\end{equation}

{\par} 
The mixture of phases with the same $n$ has 
\begin{equation}
  \label{eqGmix}
  G(n,x_i) = \sum_i x_i G_i (n) , 
\end{equation}
where $x_i$ is the $i$-phase  fraction, and 
\begin{equation}
  \label{eqxsum}
  \sum_i x_i  =1 . 
\end{equation}

{\par} 
At thermodynamic equilibrium, $G_2^{(0)} =  G_1^{(0)} \equiv G_0$. 
Neglecting the higher-order terms $O(n-n_i)^3$, we get:
\begin{equation}
  \label{eqGn}
   G(n,x_1) \approx G_0 + x_1  (n-n_1)^2 G_1^{(2)} + (1-x_1) (n-n_2)^2 G_2^{(2)} 
\end{equation}
If both phases are stable, then $G_i^{(2)} \equiv (\partial^2 G_i / \partial n^2)_{n_i} >0$. 
We can generalize consideration to any continuous  curves $G_i (n)$ with minima at $G_i (n_i) = G_i^{(0)}$; 
each curve is convex at the minimum $n_i$ and monotonic on both sides (decreases at $n<n_i$ and increases at $n>n_i$),
see Fig.~\ref{FigGn}.

{\par}
Without a loss of generality, let us assume $n_1 < n_2$ (i.e., we label the phase with lower $n$ by index 1). 
Than for any intermediate electronic density $ n_1 <n< n_2 $, 
segregation into two phases with electronic densities $n_1'$ and $n_2'$, 
where $n_1 < n_1' < n_b$ and $n_b < n_2' < n_2$, 
results in the lowering of $G$, and is favorable, see Fig.~\ref{FigGn}. 


\subsection{\label{Instability}Instability is repulsive}
{\par} 
	Obviously, electronic and lattice structure at the instability is unstable, while it becomes more stable further from the instability.  
Increased  ``distance''  form the instability
in terms of the phase space coordinates 
 results in lower $G$. 
Thus, instability is ``repulsive'' in the phase space.  
This is illustrated by Fig.~\ref{FigGn}, where $G_b(n_b)$ is the instability, and $n$ is a phase space coordinate. 
The ``repulsive'' region is at $n_1 < n < n_2$. 


\subsection{\label{SS}Superconductivity due to charged segregation}
{\par} 
	Again, one way to move away from the instability in terms of electron charge density $n$ or doping $\Delta n$ is a charge density segregation, which leads to creation of charged precipitates. 

{\par} 
	Let us consider a charged phase with a fixed total volume $V_Q$ and fixed total charge $Q$. 
If this phase is allowed to fracture to $N_q = Q/q$ small precipitates of charge $q$,  
then its 
 total potential energy $ U \sim q^{2/3} Q^{4/3} V_Q^{-1/3}$ 
will be minimal for the smallest $q$, which nevertheless cannot be smaller than a certain quantum limit. 
This leads to quantum charges $q$ of the precipitates, which can behave as quasi\-particles.
If these quasi\-particles are bosons, then they can form a Bose-Einstein condensate \cite{Bose1924} at low $T$. 
A condensate of charged bosonic quasi\-particles is responsible for superconductivity. 

{\par} 
	Charged quasi\-particles repel each other.  This repulsion distributes them uniformly (in the absence of external fields), and can order them into a ``quasi\-lattice'' (a geometric lattice-like arrangement of quasi\-particles). 

{\par} 
	Without doubt, a variation of an electronic structure and charge density causes a lattice deformation. From the other hand, a local lattice deformation causes a local variation of the electronic structure: this reminds ``the chicken and the egg'' problem. Charged precipitates are so small, that they must be coherent with the lattice, but this coherency does not prevent them from creating a local strain.  
Symmetry of $d_a$ distribution around a quasi\-particle in a superconductor
differs from a thermal distribution of inter\-atomic distances, especially at low $T$.  
This lattice deformation could be detected in experiment using diffraction of x-rays or neutrons.

\subsection{\label{Magnetism}Magnetism} 
{\par}
	The role of magnetism in the high-$T_c$ superconductors is still debatable.  Typically there are several competing magnetic orderings around the instability, and the border  between those spin states is also an instability of the electronic structure.  Our model is applicable to any electronic instability.  To remain generic, we do not restrict our consideration to a particular kind of instability, which might \cite{PhysicsToday51n10p40y1998,Nature468p283y2010} or might not \cite{PRB95p174301y2017} be magnetic.

\begin{figure}[b]
\includegraphics[width=75mm]{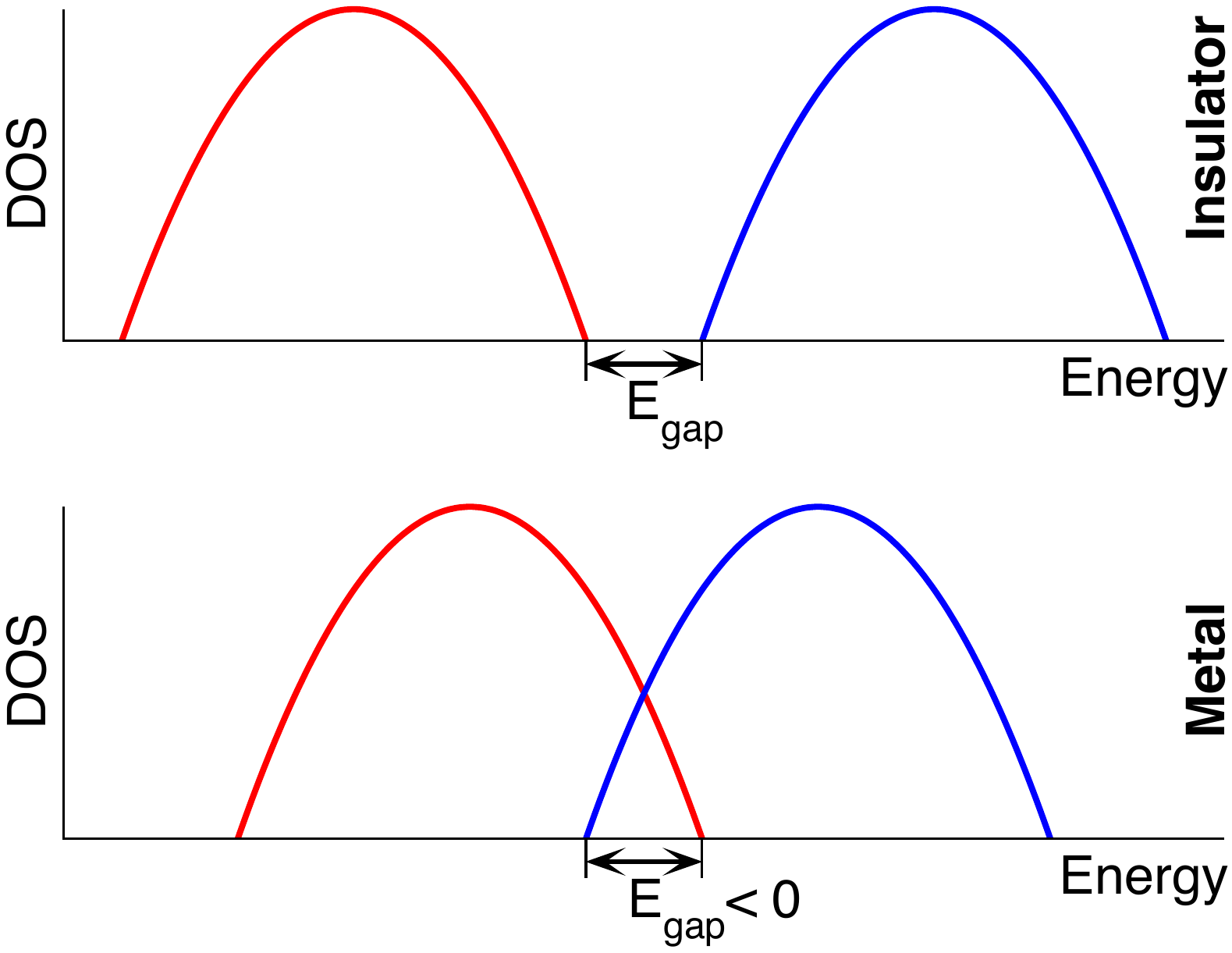}
\caption{\label{FigGap} 
Transformation between 
an insulator (with a band gap $E_{gap} \! > \! 0$) and 
a metal (with a band overlap, $E_{gap} \! < \! 0$).
}
\end{figure}

\begin{figure}[t]
\includegraphics[width=75mm]{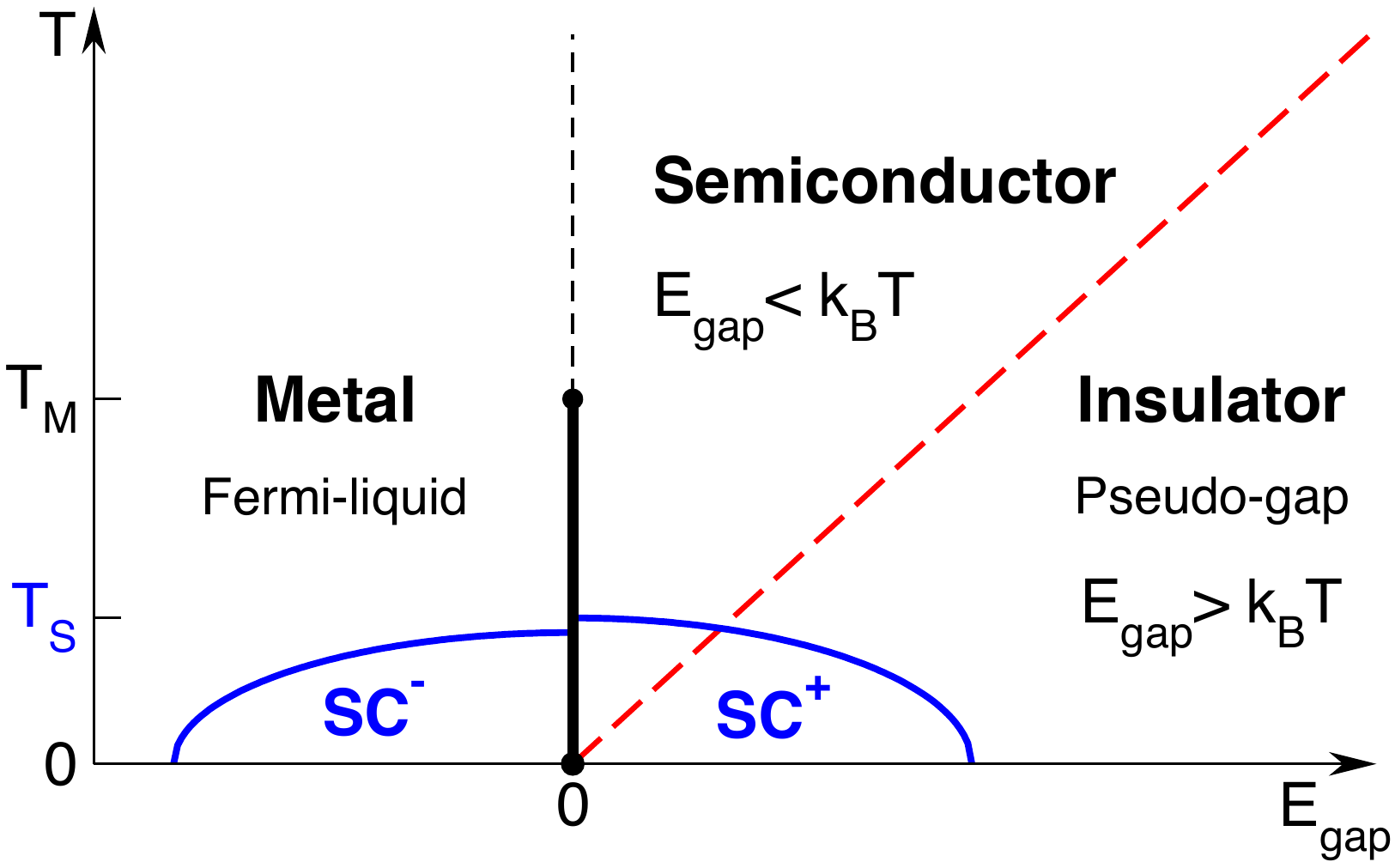}
\caption{\label{FigTE} Temperature $T$ vs. the band gap $E_{gap}$ near the Mott transition.
Mott transition is 1st-order below the critical point at $T_M$. 
Superconducting state (SC) appears at low $T \le T_S = \mbox{max}(T_S^- , T_S^+)$. 
In a metal, $E_{gap} < 0$ is an overlap of bands, which changes monotonically with a finite electronic density at the Fermi level for small overlaps.
Material with a band gap $E_{gap} >0$ is either an insulator or a semiconductor, which conducts electricity if $E_{gap}<k_B T$. }
\end{figure}

\begin{figure}[t]
\includegraphics[width=75mm]{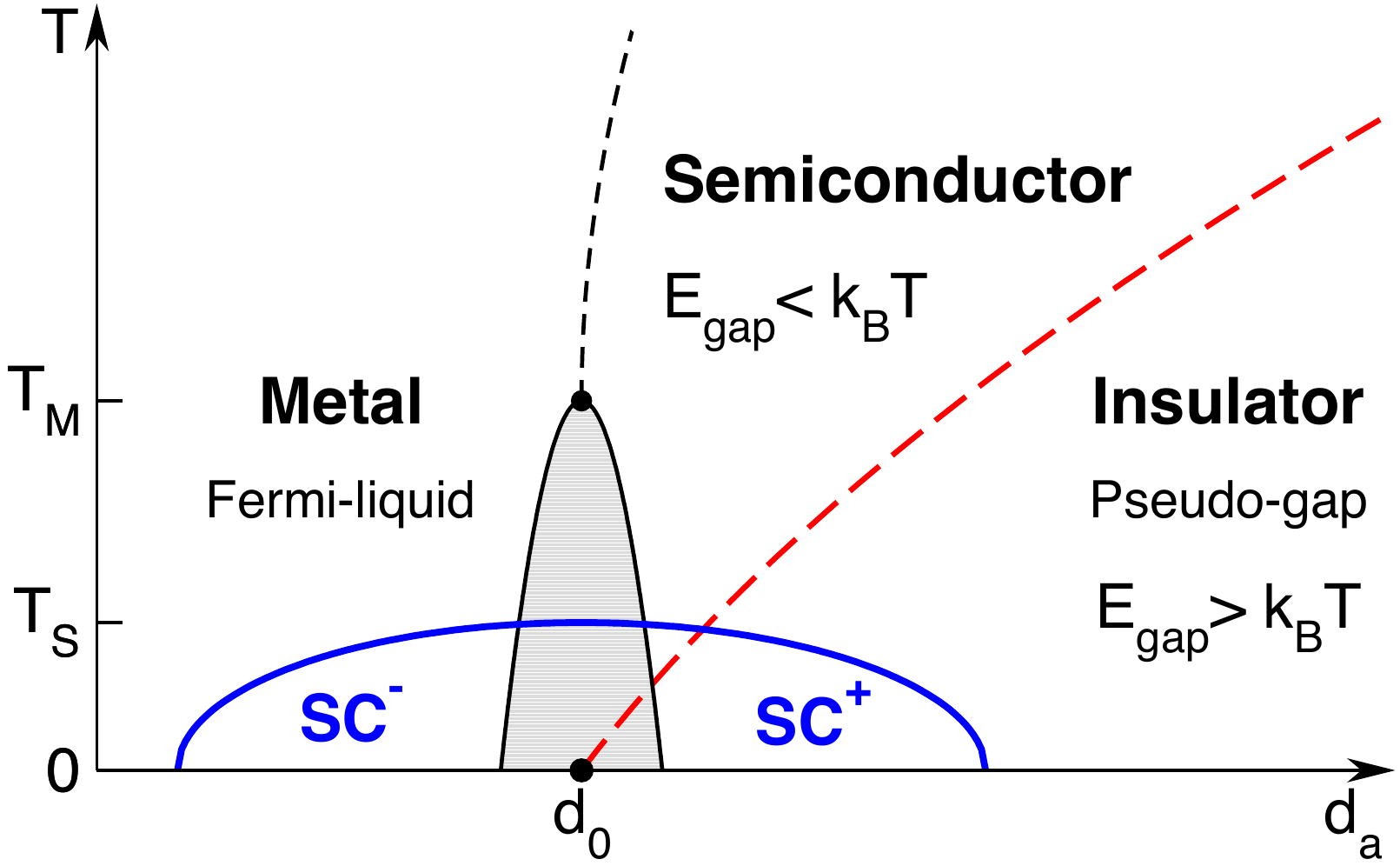}
\caption{\label{FigTd} Temperature $T$ vs. characteristic interatomic distance $d_a$  near the Mott transition.  
For two phases with different electronic structure at thermodynamic equilibrium, 
there is a gap in $d_a$, as well as lattice constants, density, and unit cell volume below $T_M$. 
The QCP at ($d_0$,$T$=0) is in this gap.
On the two sides of the gap, $T_S^-$  and $T_S^+$ can differ. 
}
\end{figure}

\section{\label{Mott}Mott transition} 
{\par} 
	An example of a phase transformation, which changes topology of the electronic structure (Fig.~\ref{FigGap}), is the Mott transition.
	Figs.~\ref{FigTE} and \ref{FigTd} show 5 distinctive phases: 
metal ($E_{gap} \! < \! 0$), a small-gap semiconductor ($0 \! < \! E_{gap} \! < \! k_B T$), insulator ($E_{gap} \! > \! k_B T$), 
and two superconducting phases at $T \! < \! T_c(E_{gap}) \! \le \! T_S$ on both sides of the instability at the phase transition at $E_{gap} \! \equiv \! 0$, which is of the first-order at $0 \! \le \! T \! \le \! T_M$.  
Again, Fig.~\ref{FigTE} shows one instability and 5 different solid phases around it.  Some of those phases can be uniform, while others (including both CS) are segregated states. 

\subsection{\label{Segregation}Neutral and Charged Segregation} 
{\par} 
	Mott transition is accompanied by both electronic and lattice instability. 
Fig.~\ref{FigTd} shows a shaded region, where crystal structures are unstable.  Such unstable structures segregate into charge-neutral stable phases of higher and lower density ($\rho$, as well as $V$ and $d_a$). 

{\par} 
In addition, the blue line in Fig.~\ref{FigTd} is the border of a region of electronic instability, where the electronic structure wants to segregate into charged regions of higher and lower electronic density $n$, and 
above we provided a model explaining  
why this segregation is energetically favorable. 
Superconductivity is a result of this charged segregation. 
Conservation of electric charge (and consequently charge neutrality of the whole system) is an additional constraint, 
imposed on charged segregation.

\subsection{\label{MottDiagrams}Comparison to other diagrams} 
{\par} 
{\par} 
	Figs.~\ref{FigTE} and \ref{FigTd} are generic phase diagram for the Mott transition.  A compatible $T$--$P$ diagram for a compressible lattice is shown in Fig.~1 in \cite{PRL109p176401y2012}, while the gap in strain and $d_a$ is shown in their Fig.~4. 
Generic $T$--$c$ diagrams in Fig.~1 in \cite{nmat8p630y2009} and Fig.~2 in \cite{PhilTransRSocA369p1574y2011} show the small-gap semiconductor (at $0 \! < \! E_{gap} \! < \! k_B T$) as a ``strange metal''.

{\par} 
	Early attempts to draw a generic diagram may contain errors. 
In particular, 
Fig.~150 in \cite{RevModPhys70p1039y1998} provides a schematic phase diagram of pseudogap structure in high-$T_c$ cuprates, 
but  it does not show a QCP at $T=0\,$K.

\section{\label{Experiment}Comparison to Experiment} 
{\par} 
	In theory, phase diagrams showing $T$ versus a phase-space variable $x$ such as the band gap $E_{gap}$ (Fig.~\ref{FigTE}), overlap of electronic orbitals, characteristic inter\-atomic distance $d_a$ (Fig~\ref{FigTd}), average electronic density $n$, or its change $\Delta n$,  
should be comparable regardless of the cause of variation of $x$.  Examples of such causes are variations of  composition $c$ or applied pressure $P$ \cite{PRL101p057006y2008}. 
Indeed, experiment \cite{NatMater8n6p471y2009} shows similarities between structural distortions
under pressure and chemical doping in superconducting BaFe$_2$As$_2$.  
Similar effects are found in SrFe$_2$As$_2$ doped by Co \cite{PSSB254n1p1600154y2017}
and CaFe$_2$As$_2$ doped by Sr \cite{PRB94p144513y2016}.

{\par} 
Examples of experimental $T$--$c$ phase diagrams are 
Fig.~6 in \cite{PRB79p014506y2009} for the electron-doped 
Ba(Fe$_{1-x}$Co$_{x}$)$_{2}$As$_{2}$,
Fig.~1 in \cite{PRB93p094513y2016} for Ba$_{1-x}$Rb$_x$Fe$_2$As$_2$,
Fig.~4 in \cite{Science329p824y2010} for Ba(Fe$_{1-x}$Co$_x$)$_2$As$_2$, and
Fig.~1d in \cite{NMat7p953y2008} for CeFeAsO$_{1- x}$F$_x$.
Examples of experimental $T$--$P$ diagrams are 
Fig.~3 in \cite{SRep4p3685y2014} for SrFe$_2$As$_2$, or
Fig.~3 in \cite{nmat8p630y2009} for Fe$_{1.01}$Se.  
Structural instability and superconductivity in SrNi$_2$(P$_{1−x}$Ge$_x$)$_2$ solid solutions was studied in \cite{PSSB254n1p1600351y2017}.  
Magnetic and structural transitions of SrFe$_2$As$_2$ at high $P$ were investigated in \cite{SRep4p3685y2014}.  
Experiment finds that superconductivity happens around an instability,
and theory claims  that it happens due to  an instability. 

\section{\label{Discussion}Discussion} 
\subsection{\label{SC2}Two types of superconductivity}  
{\par}
	Our model of a charged segregation predicts that charges $q$ of tiny precipitates (quasi\-particles) should be quantized (e.g., $q=2e$ for the Cooper pair), 
but it allows both signs of $q$.  Hence, we anticipate existence of two distinctive types of superconductivity with positive and negative $q$.  
Here one can find  similarity to two types semiconductors: p- and n-type.  However, charge carriers in semiconductors can be fermions, while in a SC they are bosons. 
Two distinctive types of superconductivity are labeled $SC^-$ and $SC^+$ in Figs.~\ref{FigTE} and \ref{FigTd}.

\subsection{\label{Collective}Collective excitations} 
{\par}
	We mentioned that electrons move much faster than their collective excitations.   
In particular, in a metal the Fermi velocity of electrons $v_F \sim 10^6\,$m/s is huge compared to the drift velocity of charge carriers $u_D \ll 10^{-3}\,$m/s. 
Thus, all quasi\-particles in a superconductor (including the Cooper pair) are collective electronic excitations, which should not be confused with 
propagation of a pair of particular electrons.  Electrons in this collective excitation change, but the charge of the excitation and its total spin (responsible for bosonic behavior) remain constant. 
{\par}
	Each quasi\-particle is a collective electronic excitation (which locally changes the charge density), accompanied by a lattice deformation.  
Its motion with a small drift velocity $u_D$ is accompanied by equally slow motion of that lattice deformation.  Particular atoms or ions vibrate around their lattice positions; they do not follow the quasi\-particle. 
However, a local change in density can be positive or negative; it is responsible for the mass of a quasi\-particle. 
{\par} 
	There is a coupling between a collective electronic excitation and a collective atomic displacement.   An electron-phonon coupling is one type of such coupling, but not the only one.

\subsection{\label{Phonons}Lattice deformations and Phonons}
{\par}
	Conventional superconductivity \cite{Onnes1913,Abrikosov2003,JETP35p1558y1959} happens due to electron-phonon coupling \cite{GinsburgLandau1950,PR106p162p1175y1957,JETP34p58p73y1958,JETP34p66y1958,PRB93p054517y2016}.
However, not every lattice deformation can be explained in terms of phonons. 
In particular, atomic positions in one phase are not always related to phonons in another phase in a phase-segregated material. 
Next, phonons are (quasi)harmonic vibrations of atoms, but not every collective atomic motion in a solid is harmonic. 
Hence, there are several reasons, why conventional theory of superconductivity might fail in several classes of  ``unconventional'' superconductors. 
{\par}
	From the other hand, our description of a superconductor as a phase segregated material with charged segregation 
is applicable to both conventional and high-$T_c$ superconductors.

\section{\label{Summary}Summary} 
{\par} 
	
We proposed a qualitative model of superconductivity, based on thermodynamics of a charged phase segregation.  
We described a superconductor as a segregated material with the quantized charge of tiny precipitates, 
which behave as charged bosonic quasi\-particles.  
A Cooper pair was mentioned as an example of such quasi\-particle. 
With cautions, our model can be viewed as a generalization of the conventional theory of superconductivity 
\cite{JETP35p1558y1959,GinsburgLandau1950,PR106p162p1175y1957,JETP34p58p73y1958,JETP34p66y1958}. 

{\par} 
We pointed at instability of the electronic structure and the lattice as a cause for phase segregation.  
As an example, we considered instability at the Mott transition,   
around which we labeled 5 distinctive solid phases (shown in Figs.~\ref{FigTE} and \ref{FigTd}), 
two of which are superconductive.  

{\par} 
We linked superconductivity with both the instability of the electronic structure and the lattice response to variations of charge density.  
We claimed that a superconductor with a higher $T_c$ has  
a larger lattice response, which can stabilize the charged bosonic quasi\-particles at higher $T$. 
Thus, our model can be used in a guided search for novel high-$T_c$ superconductors.

\acknowledgments
\begin{acknowledgments}
We acknowledge Paul C. Canfield and Duane D. Johnson for discussion.
This work was supported in part by the U.S. Department of Energy (DOE), Office of Science, Basic Energy Sciences, Materials Science and Engineering Division. The research was performed at the Ames Laboratory, which is operated for the U.S. DOE by Iowa State University under contract \# DE-AC02-07CH11358.
\end{acknowledgments}

\bibliography{Mott}

\end{document}